\documentclass[10pt,conference]{IEEEtran}
\usepackage{amsmath, amssymb, amsthm, graphicx, cite}

\newtheorem{thm}{Theorem}

\newtheorem{lem}{Lemma}

\IEEEoverridecommandlockouts 

\title{Communication Strategies for Low-Latency Trading}
\author{
\IEEEauthorblockN{Mina Karzand}
\IEEEauthorblockA{Massachusetts Institute of Technology}\and
\IEEEauthorblockN{Lav R.\ Varshney}
\IEEEauthorblockA{University of Illinois at Urbana-Champaign}
}

\begin{document}

\maketitle

\begin{abstract}
The possibility of latency arbitrage in financial
markets has led to the deployment of high-speed communication
links between distant financial centers. These links are noisy and
so there is a need for coding. In this paper, we develop a game-theoretic
model of trading behavior where two traders compete to
capture latency arbitrage opportunities using binary signalling. Different coding schemes
are strategies that trade off between reliability and latency.
When one trader has a better channel, the second trader should not compete. 
With statistically identical channels, we find there are two different
regimes of channel noise for which: there is a unique Nash equilibrium
yielding ties; and there are two Nash equilibria with different winners.  
\end{abstract}

\vspace{2mm}
{\small\emph{``One guy says to me, `It doesn't matter if I'm one second slower or one microsecond; 
either way I come in second place.'''}} \\
\hspace*{\fill} {\small--- Flash Boys: A Wall Street Revolt \cite[p.~63]{Lewis2014}} 

\section{Introduction}
\label{sec:introduction}

The traditional view in information theory is that communication reliability is to be 
obtained by paying a penalty in latency.  Though there is growing interest in
finite blocklength information theory \cite{PolyanskiyPV2010,ChenLM2013,Tan2014}, motivated by general notions
that latency is problematic, few works have put forth
explicit reasons for needing to meet latency constraints, cf.~\cite{VarshneyMG2012}.

When there are latency arbitrage opportunities in financial markets, however,
latency is a key performance metric \cite{Aldridge2013}.  One must be faster than one's 
competitors.  Indeed, the speed of light is a key constraint
and physical distance a key consideration when building communication channels, whether fiber optic
or microwave \cite{MacKenzieBMP2012,Lewis2014}.  

When prices of the same financial instrument in different financial markets (say Chicago and New York) diverge for a short period 
of time, this is called a latency arbitrage opportunity.  Low-latency traders jump in and trade away the price discrepancies.  
As an example, a low-latency trader would sell pork bellies in the market where the commodity is temporarily overpriced, while simultaneously 
buying it in the distant market where the commodity trades too cheaply.  In the process, the demand and supply produced by the low-latency trader 
equilibrates market prices in markets that were previously divergent.  Then, the low-latency trader quickly reverses his position to capture the gain, 
and investors of all frequencies can be assured that prices on traded financial instruments are consistent across the globe, upholding what is called 
the \emph{law of one price} \cite{Aldridge2013}.

What are the communication requirements for a latency arbitrageur?  Some unreliability is allowable since there is not a need 
for perfect information to make riskless profits; there is only need to skew the odds systematically in one's favor \cite[p.~74]{Lewis2014}.
A recent article noted \cite{Schneider2012}: 
\begin{quotation}
\noindent that high-frequency traders would
much rather have access to a communications channel that's
faster than every one else's, even if it gets flaky every now and
then. A link that's second or third fastest isn't of much use to
them, even if it's always available. That's a very different calculus
than the one most engineers use--but it's clearly the one
you want to follow if you're trying to get ahead of the pack.
\end{quotation}
Designing for such novel requirements suggests new problems at the intersection 
of communication theory and game theory, and provides a concrete
reason to step back from infinity.  Some previous work on low-latency communication
did not consider the competitive nature of the communication \cite{Maric2013}.

One should be careful not to confuse low-latency trading with ``high-frequency trading.''  Low-latency refers to the ability to quickly route and 
execute orders irrespective of their position-holding time, whereas high-frequency refers to the fast turnover of capital that may require low-latency 
execution capability \cite{Aldridge2013}.  Although academic study of low-latency trading has been limited, 
Moallemi and Sa\u{g}lam put forth three main reasons for wanting low latency
in trading \cite{MoallemiS2013}:
\begin{enumerate}
  \item \emph{Staleness}: A trader with significant latency will make trading decisions 
	based on information that is stale.
  \item \emph{Relative Latency}: A trader will want to act to exploit a 
	discrepancy before a price correction takes place, i.e.\ before competitors are able to act.
  \item \emph{Ordering}: Modern markets treat orders differentially based on time of arrival.
\end{enumerate}
They focus on a model that captures staleness, but do not consider relative latency.
The goal herein is to look at relative latency through a mathematical model of repetitive binary signalling
that is essentially a decoding game, where decoding time is the strategy.  

Our main results show that when one trader has a better channel, the second trader should not compete. 
With statistically identical channels, we find there are two different
regimes of channel noise for which: there is a unique Nash equilibrium
yielding ties; and there are two Nash equilibria with different winners, respectively.
A question commonly asked by market participants and regulators alike is how much speed is enough?  
Clearly, the race for speed will end when there is equilibrium: when an additional dollar spent on technology no longer generates 
extra return \cite{Aldridge2013}.  Our results also take steps to answering this question.
  
\section{Problem Formulation}
We put forth a stylized model of low-latency trading.

\subsection{Arbitrage Payoffs}
Let us assume there are two firms A and B who want to exploit latency arbitrage between markets I and II. Both firms have agents in both markets. There
is a single commodity under consideration and the values of trades are fixed.  Hence there are just two types of messages: 
``Buy'' and ``Sell''.\footnote{In the real system, messages are transmitted via a Financial Information eXchange (FIX) protocol.
A typical FIX message is composed of a header, a body containing order execution directives, and a checksum \cite{Aldridge2010}.}  
The message is transmitted by the agent of each firm in market I. The agents in market II try to recover their message correctly with high 
probability and as fast as possible. Payoffs are determined by the order in which the firms act. So we describe payoffs accordingly.
\begin{itemize}
\item \emph{First Firm}: The first firm to act in market II based on its estimate of the message has the following payoffs:
\begin{itemize}
\item If the transmitted message is ``Buy,'' and the agent in market II correctly recovers it, the firm earns $V_1$ dollars. 
If the agent makes a mistake and assumes the transmitted message was a ``Sell,'' the firm loses $V_1$ dollars (i.e., the payoff is $-V_1$ dollars).
\item If the transmitted message is ``Sell,'' and the agent in market II correctly recovers it, the firm earns $V_2$ dollars. 
If the agent makes a mistake and assumes the transmitted message was a ``Buy,'' the firm loses $V_2$ dollars (i.e., the payoff is $-V_2$ dollars).
\end{itemize}

\item \emph{Second Firm}: Since we assume the market has equilibrated after the first transaction, the second firm decoding the message will have zero payoff 
whether or not it recovers the message correctly.

\item \emph{Simultaneous Decoding}: If both firms perform decoding and act upon it at the same time, each of them receives half of the payoff that the first firm would receive.

\item \emph{Efficient Market}: We assume that after time $T_0$, the market becomes efficient through the action of high-latency traders and the arbitrage opportunity disappears. 
\end{itemize}
We assume the prior probabilities $P_1$ and $P_2$ for the messages ``Buy" and ``Sell," where $P_1+P_2=1$.  Furthermore, each transaction has a cost for any firm
since there are real costs associated with accessing an exchange, ranging from payment for direct market feeds to managing routing logic: we assume a cost of 
$c$ dollars per transaction. Further, we assume that each firm pays $dS$ dollars to use a channel with signal-to-noise ratio $S$;
this is a linear function of signal power as in most physical communication transmitters. The maximum signal-to-noise ratio of the channel is $S_0$.

Clearly, this is a first step towards modeling the payoffs in high-speed trading in financial markets. There are other possible ways to model payoffs depending on the market under consideration. 

\subsection{Strategy Space}

The strategy space of each firm is a nonnegative integer, denoted by $T_A$ and $T_B$, respectively which is the delay
in before decoding the message.  Specifically, the decision time of firm A, $T_A$, and the decision time of firm B, $T_B$
are the times when the agents in market II of each firm decide to decode the message and act based on 
their estimates of the transmitted message.  The fundamental speed-accuracy tradeoff can be parameterized by $T$: the larger the decoding delay 
for a firm, the smaller probability of error it has on average. Thus, the expected payoff would be higher, \emph{if it is the first firm to decode and act}. 
But as the delay increases, it gets less likely for a firm to be the first actor.

\subsection{Communication Scheme}
For concreteness, let us suppose that we are communicating over a discrete-time
AWGN channel with power constraint $P$.  The noise power is assumed to be $N_0$, so that the signal-to-noise ratio of the channel is 
$S=P/N_0$.  

We assume that each agent in market I uses binary phase-shift keying (BPSK) repetition coding. Admittedly, this is a very simple model of the communication channel and coding scheme. But we believe that studying this model captures the essence of the latency-accuracy tradeoff that is of interest to us.   If the message is ``Buy,'' the agent sends 
$+a$ into the AWGN channel; the received symbol is a normally-distributed random variable $\mathcal{N}(+a,N_0)$. Similarly, if the message 
is ``Sell,'' the agent sends $-a$ into the AWGN channel; the received symbol is a normally-distributed random variable $\mathcal{N}(-a,N_0)$.
The power constraint implies the optimal choice $a=\sqrt{P}$.

The agent at market II estimates the transmitted message after receiving $T$ symbols, $y_1,\dots,y_T$. The sufficient statistic of the received 
symbols is $\frac{1}{T}\sum_{i=1}^{T}{y_i}$. The decoder compares this sufficient statistic with a chosen threshold, $h$. The optimal method to choose the 
threshold will be discussed later. Thus, optimal decoding is:
\[
\frac{1}{T}\sum_{i=1}^{T}{y_i} \quad\mathop{\gtreqless}_{\text{Sell}
}^{\text{Buy}}\quad h \sqrt{P}
\]

There are two types of error events:
\begin{itemize}
\item If the transmitted message was ``Buy'' and the decoder decided ``Sell,'' we have the error event of type 1. The probability of this error event is:
\begin{align*}
	P_{e,1} & = \mathbb{P}(\text{error}|\text{Buy})\\
	& = \mathbb{P}(\frac{1}{T}\sum_{i=1}^{T}{y_i}<h\sqrt{P} | \text{Buy})\\
	& = Q(T \frac{a-h\sqrt{P}}{\sqrt{T N_0}}) = Q(\sqrt{T \frac{P}{N_0}(1-h)^2})\\
	& = Q(\sqrt{2 T S \gamma_1}) \approx \exp(-TS\gamma_1)
\end{align*}
where $Q(\cdot)$ is the cumulative distribution function of the standard normal distribution, and 
\begin{equation}
\label{eq:gamma1}
\gamma_1=(1-h)^2/2.
\end{equation}

\item If the transmitted message was ``Sell'' and the decoder decided ``Buy,'' we have the error event of type 2. The probability of this error event is:
\begin{align*}
	P_{e,2} & = \mathbb{P}(\text{error} |  \text{Sell})\\
	& = \mathbb{P}(\frac{1}{T}\sum_{i=1}^{T}{y_i}>h \sqrt{P} | \text{Sell})\\
	& = Q(T \frac{a+h\sqrt{P}}{\sqrt{T N_0}})= Q(\sqrt{T \frac{P}{N_0}(1+h)^2})\\
	& = Q(\sqrt{2T S\gamma_2}) \approx \exp(-TS\gamma_2)
\end{align*}
where 
\begin{equation}
\label{eq:gamma2}
\gamma_2=(1+h)^2/2.
\end{equation}
\end{itemize}

\subsection{Expected Payoff}

\begin{itemize}
	\item \emph{Expected payoff of the first firm:} 
	As mentioned, with probability $P_1$ the transmitted message is ``Buy" and with probability $P_2$, it is ``Sell". The first firm decodes the transmitted message 
	based on its own estimation of the message and makes a transaction (and therefore pays the transaction fee of $c$ dollars). Thus, the expected payoff of the first 
	firm is as following:
	\begin{align*}
		&\mathbb{E}[\text{payoff of first firm}] \\
& \ =  \mathbb{E}[\text{payoff}|\text{being first}]-c- d S\\
		& \ =  P_1 \mathbb{E}[\text{payoff}|\text{message is Buy and }\text{first}] \\
& \ \qquad+ P_2 \mathbb{E}[\text{payoff}|\text{message is Sell and }\text{first}]-c-d S\\
		&  \ =  P_1 \left[(1-P_{e,1})V_1 + P_{e,1}(-V_1)\right] \\
& \ \qquad + P_2 \left[(1-P_{e,2})V_2 + P_{e,2}(-V_2)\right]-c - d S\\
		& \ =  P_1 V_1 (1 - 2P_{e,1})+ P_2 V_2 (1 - 2P_{e,2})-c - d S\\
		& \ \approx P_1 V_1 [1 - 2e^{-TS\gamma_1}]  + P_2 V_2 [1 - 2e^{-TS\gamma_2}]-c - d S
	\end{align*}
		
	\item \emph{Expected payoff of the second firm:}
		The expected payoff of the second firm is $-d S$ regardless of its decoding probability of error, since the payoff of the transaction is zero.
		The second firm does not pay the transaction fee as it does not make any transaction.
\end{itemize}

If the decision time of firms are $T_A = T$ and $T_B$, and also the signal-to-noise ratios are $2S\gamma_{1,A}$, $2S\gamma_{2,A}$, $2S\gamma_{1,B}$ and $2S\gamma_{2,B}$, then the expected payoff of firm A would be:
\begin{align*}
	&\pi_{A}(T,T_B)  \\
&= \mathbb{I}\{T<\min\{T_B,T_0\}\}\left[P_1 V_1 [1 - 2e^{-T S_A\gamma_{1,A}}]\right.\\
&\qquad+ \left.P_2 V_2 [1 - 2e^{-T S_A\gamma_{2,A}}]-c- d S_A\right]\\
&\quad+ \mathbb{I}\{T=\min\{T_B,T_0\}\}\left[P_1 V_1 [1 - 2e^{-T S_A\gamma_{1,A}}]/2\right.\\
&\qquad+ \left.P_2 V_2 [1 - 2e^{-T S_A\gamma_{2,A}}]/2-c- d S_A\right]\\
&\quad+ \mathbb{I}\{T > \min\{T_B,T_0\}\}\left[- d S_A\right]
	\end{align*}
and symmetrically for the other firm.

\section{Optimizations}
Having defined the players, the strategies, and the payoffs in this standard form game, we can now
consider some optimization problems, before turning to questions of equilibria.

\subsection{Optimal Decoding}
We study an optimal adaptive decoding strategy in which the threshold value and tradeoff point between two error types of the decoder 
depends on its chosen value of delay. 

Let the decoder choose its decoding threshold knowing the decision time:
\begin{align*}
	h^*_A(T_A,S_A) & = \arg\max_{h} P_1 V_1 [1 - 2e^{-T_A S_A\gamma_{1,A}}] \\
&\quad\quad +P_2 V_2 [1 - 2e^{-T_A S_A\gamma_{2,A}}] \\
	& \approx \frac{P_1 V_1 - P_2 V_2}{P_1 V_1 + (1-2T_A S_A) P_2 V_2}		
\end{align*}
Note that this choice of threshold captures the tradeoff appearing due to different expected payoff of transmitted messages. 
The decoder can reduce the probability of error given the message was a ``Buy'' at the cost of a higher probability of error for the ``Sell'' messages.
\[
\gamma^*_1(T)=(1-h^*(T,S^*))^2/2
\]
\[
\gamma^*_2(T)=(1+h^*(T,S^*))^2/2.
\]
We will also need to define the following functions for use in equilibrium analysis:
\begin{align*}
&F_A(T_A)\\
&=P_1 V_1 (1 - 2P_{e,1,A})+ P_2 V_2 (1 - 2P_{e,2,A})\\
&\approx P_1 V_1 [1 - 2e^{-T_A S^*_A\gamma^*_{1,A}(T_A)}]+ P_2 V_2 [1 - 2e^{-T_A S_A\gamma^*_{2,A}(T_A)}]
\end{align*}
\begin{align*}
&F_B(T_B)\\
&=P_1 V_1 (1 - 2P_{e,1,B})+ P_2 V_2 (1 - 2P_{e,2,B})\\
& \approx P_1 V_1 [1 - 2e^{-T_B S^*_B\gamma^*_{1,B}(T_B)}]+ P_2 V_2 [1 - 2e^{-T_B S_B\gamma^*_{2,B}(T_B)}]
\end{align*}

\subsection{Optimal Power Allocation}
The transmitter needs to determine the allocated signal-to-noise ratio ahead of time before the communication takes place,
since this is typically not adaptable at short timescales.  We further assume the signal-to-noise ratio is determined not knowing the actual 
decision time. Thus, the allocated signal-to-noise ratio is assumed to be optimal for a chosen decision time. We assume that the chosen decision time 
which determines the signal-to-noise ratio is the time market becomes efficient, $T_0$.
\begin{align*}
S^*&=\arg\max_{S\leq S_0} P_1 V_1 [1 - 2e^{-T_0 S\gamma_{1}}]/2\\
&\quad\quad\quad\quad\quad+ P_2 V_2 [1 - 2e^{-T S\gamma_{2}}]/2-c- d S
\end{align*}
where 
\[h^*=\frac{P_1 V_1 - P_2 V_2}{P_1 V_1 + (1-2T_0 S) P_2 V_2}\] 

and $\gamma_1$ and $\gamma_2$ are defined as in~\eqref{eq:gamma1} and~\eqref{eq:gamma2}.

\section{Equilibria}
Now we consider Nash equilibria for the game.  We observe that the expected payoff of firm A when it chooses the optimal decoder is the following:
\begin{align}
	\pi_{A}(T_A,T_B)&=  \mathbb{I}\{T_A<\min\{T_B,T_0\}\}\left[F_A(T_A)-c-d S_A\right] \nonumber \\
	&\quad +  \mathbb{I}\{T_A=\min\{T_B,T_0\}\}\left[F_A(T_A)/2-c-d S_A\right] \nonumber \\
	&\quad + \mathbb{I}\{T_A>\min\{T_B,T_0\}\}\left[-d S_A\right]
\label{eq:expPayoff}
\end{align}

Since $F_A(T_A)$ is a strictly increasing function, given $T_B$ and $T_0$, the optimal strategy for firm A is as follows:

\begin{equation}
	\label{eq:response}
	T^*_{A}(T_B)=\begin{cases} 
      T_B+1, &  c \geq \max\{F_{A}(T_B-1),F_{A}(T_B)/2\}\\
      T_B-1, &  F_{A}(T_B-1)\geq\max\{c,F_{A}(T_B)/2\}\\
      T_B, & F_{A}(T_B)/2\geq \max\{F_{A}(T_B-1),c\}
   \end{cases}
\end{equation}

Similarly for firm B. Now we try to find the equilibrium of this game in the symmetric case. 
In this scenario, the functions $F_{A}(\cdot)$ and $F_{B}(\cdot)$ behave similarly. 

We are interested in identifying the long-term behavior of the system when this game is repeated many times. Thus, we assume the agents 
follow the best response dynamics defined as follows: the game is repeated many times and each firm at each repetition of the game chooses 
the best strategy based on the strategy of the competitor in the previous repetition. Conditioned on the monotonicity of the payoff function, 
when there exists a Nash equilibrium in the game, the best response dynamics converges \cite{FudenbergT1991}. 

We observe we can expect two types of behavior from this game depending on the regime in which parameters are defined:
\begin{enumerate}
		\item[TIE] For some $T^*$, there is Nash equilibrium $(T_A,T_B)=(T^*, T^*)$ and $(T_A,T_B)=(T^*, T^*)$. This is a valid Nash equilibrium of the game if the following properties hold:
	\begin{equation}
		\label{eq:eqNash}
			F(T^*)/2 \geq \max\{F(T^*-1),c\}
			\end{equation}
			
	\item[WIN] For some $T^*$, we have two Nash equilibria $(T_A,T_B)=(T^*+1, T^*)$ and $(T_A,T_B)=(T^*, T^*+1)$. This is a valid Nash equilibrium of the game if the following properties hold:
\begin{equation}
	\label{eq:NeqNash}
		\begin{cases}
c \leq F(T^*) \leq 2c\\
F(T^* +1) \leq 2F(T^*)\\
F(T^*-1)\leq c
		\end{cases}
		\end{equation}
	\end{enumerate}
We first show that the TIE conditions and the WIN conditions partition the possibilities.  Note that the monotonicity condition of the lemma is clearly
true under optimal decoding: more observations lead to lower error probabilities and higher expected payoffs.
\begin{lem}
Let us assume function $F(\cdot)$ is strictly increasing and $F(0)<c< F(T_0)$. Then there exists a $0<T^*<T_0$ that satisfies either the condition set given in equation~\eqref{eq:eqNash} or the condition set given in equation~\eqref{eq:NeqNash}.
\end{lem}
\begin{IEEEproof}
Let us look at $T^*$ such that $F(T^*-1)\leq c < F(T^*)$. We can observe simply that if $F(T^*)>2c$, then $F(T^*)/2 \geq \max\{F(T^*-1),c\}$. Similarly if $F(T^* +1) > 2F(T^*)$, then $F(T^*+1)/2 \geq \max\{F(T^*),c\}$

\end{IEEEproof}
Now we prove the equilibrium properties under the TIE and WIN conditions, respectively.
\begin{thm}
Suppose that both agents start the game for some $(T_A,T_B)$, the repeated game under the best response dynamics converges to one of the possible Nash equilibria. 
If the set of conditions given in equation~\eqref{eq:eqNash} is satisfied for some $T^*$, the agents converge to simultaneous decoding at time $T^*$. If the set of conditions 
given in equation~\eqref{eq:NeqNash} is satisfied for some $T^*$, depending on the starting point, they converge to $(T_A,T_B)=(T^*,T^*+1)$ or $(T_A,T_B)=(T^*+1,T^*)$. 
\end{thm}
\begin{IEEEproof}
We can construct the best response state graph of the game. Each state in this graph corresponds to one possible strategy set. The transitions correspond to 
better/best responses of any one of the players to the strategy given in the previous state (in each transition, we assume that only one player can change its strategy). 
The transitions could be determined from the equations given in~\eqref{eq:response}. The pure state Nash equilibria of a game is the sink states of this state graph 
as no player can unilaterally improve his payoff in these states. 
the sink states of this state graph. 
We could have two possible types of state graphs depending which of these equations  ~\eqref{eq:eqNash} and ~\eqref{eq:NeqNash} are satisfied.   Due to the 
monotonicity of the function $F(\cdot)$, in any case this state graph is acyclic and starting from any state, it converges to one of the sink states. 
\end{IEEEproof}

Notice that the TIE conditions specify that the payoff increases faster than exponentially as a function of time, 
whereas the WIN conditions specify that the payoff increases slower than exponentially as a function of time.  Since the quality of the channel determines the rate at which the error probability decays/ payoff increases, it seems that the stronger channels would impose the TIE result between the agents, whereas the weaker channels would impose inequality in the results of the game.
is possible to win a trading race (in equilibrium) when the channel is not so noisy.

\section{Non-identical Channels}
Thus far we have been discussing games where the noise powers for the two players are identically $N_0$.
For completeness, let us state what happens when the channels do not have identical statistics.
If different firms have channels of different quality, the signal-to-noise ratio and the optimal transmitted power 
would be different for each firm. The expected payoff of each firm still follows the equation~\eqref{eq:expPayoff}.
The Nash equilibrium of this game will be asymmetric in the following sense.  Let us assume that $F_B^{-1}(c) <
F_A^{-1}(c)$. The equilibrium will be $(T_A, T_B) = (\lfloor F_A^{-1}(c)\rfloor+1,\lfloor F_A^{-1}(c)\rfloor)$. This equilibrium is intuitive as we
expect the firm with the better (or cheaper) communication channel to be more powerful and exploit the arbitrage in
the market.  This is what is observed with an incredibly expensive arms race to build better physical communication channels.

\section{Conclusion}
We have formulated a stylized model of the communication race that forms the heart of 
low-latency trading to take advantage of latency arbitrage opportunities in financial markets.
We found the existence of Nash equilibrium communication strategies: for one set of channel parameters,
there is a unique equilibrium that corresponds to ties where the two firms share the arbitrage opportunity.  
For another set of channel parameters, there are two possible equilibrium when one or the other firm wins
the opportunity.  

Our modeling approach cast time as discrete, but one might wonder what happens
in continuous-time models of communication races.  Rather than settling into Nash equilibria,
best response dynamics may go into a limit cycle since it is possible to \emph{just} act more quickly
than one's opponent; the discontinuity prevents equilibria for the real-valued strategy space.  
Even within a discrete-time model that comes from continuous time, the relationship between clocking speed 
and noise could make clocking a part of the strategy space.

Herein we have used various simplifications that can be relaxed in future work.  We can 
consider markets with more than two messages; advanced methods in coding theory would then become important
rather than simply having BPSK repetition.  Moreover, microwave links that are now deployed incur fading in addition 
to noise; the role of outage in trading could be intriguing.  Considering the game with many competitive firms rather 
than just two is another possible extension to the basic framework.

Finally, we have focused here on expected payoffs, as is typical in game theory,
but in many financial settings, risk is also a strong consideration.  By considering risk in addition to expected payoff as part of 
performance criteria, new optimizations and equilibria may arise.

\section*{Acknowledgment}
Thanks to S.\ Borade and J.~Z.\ Sun for helpful discussions. 

\bibliographystyle{IEEEtran} 
\bibliography{conf_abrv,abrv,lrv_lib}

\end{document}